# Simultaneous compression and characterization of ultrashort laser pulses using chirped mirrors and glass wedges

Miguel Miranda,[1,2,*] Thomas Fordell,[2] Cord Arnold,[2]
Anne L'Huillier,[2] and Helder Crespo[1]

[1]*IFIMUP-IN and Departamento de Física e Astronomia, Universidade do Porto,
Rua do Campo Alegre 687, 4169-007 Porto, Portugal*
[2]*Department of Physics, Lund University, P.O. Box 118,
SE-221 00 Lund, Sweden*
[*]*mmiranda@fc.up.pt*

**Abstract:** We present a simple and robust technique to retrieve the phase of ultrashort laser pulses, based on a chirped mirror and glass wedges compressor. It uses the compression system itself as a diagnostic tool, thereby making unnecessary the use of complementary diagnostic tools. We used this technique to compress and characterize 7.1 fs laser pulses from an ultrafast laser oscillator.

**1. Introduction**

The characterization of ultrashort laser pulses is often as important as the generation process itself. Since no methods exist for the direct measurement of such short events, self-referencing techniques are usually employed.

Traditionally, ultrashort pulses have been characterized by nonlinear autocorrelation diagnostics (see, e.g., [1]), which are still widely used in many laboratories. Although relatively simple to implement, these fail to provide complete information about the pulses. Still, several methods have been devised allowing for the reconstruction of the amplitude and phase of the pulses by combination of autocorrelation and spectral measurements (see, e.g. [2-4]). An important improvement over these techniques came in 1993 with the introduction of frequency resolved optical gating (FROG) [5,6]: by spectrally resolving an autocorrelation (or cross-correlation) signal, a sonogram-like trace is created from which complete characterization of a given pulse can be performed using an iterative algorithm. The quality of the retrieval is reflected by the corresponding FROG error, and the time and frequency marginals of the trace also provide a means to cross-check the results. There are many variants of FROG today, which all rely on spectrally resolving some time-gated signal. Other methods widely used today are related to the technique of spectral phase interferometry for direct electric-field reconstruction (SPIDER), first introduced in 1998 [7]. These methods do not rely on temporal gating, but instead on interferometry in the spectral domain: the spectrum of a given pulse is made to interfere with a frequency-shifted (sheared) replica of itself, and the resulting spectral interferogram is recorded. Although usually more complicated to set up, retrieving the spectral phase from a SPIDER trace is numerically much simpler than in FROG. Standard SPIDER however is very alignment sensitive and this can easily affect the measured pulse, as there is no straightforward means to determine the quality of the phase measurement. Recent SPIDER-related methods have been devised that allow overcoming this issue [8,9].

Recently, a new paradigm in pulse characterization based on phase scanning, multiphoton intrapulse interference phase scan (MIIPS) [10-12], was introduced. It consists in applying well-known spectral phases to the pulse to be characterized and measuring the resulting second-harmonic generation (SHG) signal. By finding which locally introduced amount of group delay dispersion (GDD) results in compression at a given wavelength, the original GDD of the pulse can be found, thereby allowing for the reconstruction of the unknown phase.

In all of the above techniques, the characterization of few-cycle laser pulses is still challenging and usually requires specific adaptations and materials in order to accommodate the associated broad bandwidths.

Our method is related to the MIIPS technique in the sense that a phase scan is performed on the pulse to be measured; however both the experimental setup and the phase retrieval method are substantially different, and these will provide major advantages with respect to other methods. In fact, our technique can be implemented using a standard chirped mirror compressor setup: we use chirped mirrors to ensure that the pulse becomes negatively chirped, and then add glass continuously until the pulse becomes as short as possible. Measuring the generated SHG spectra around this optimal glass insertion allows us to fully retrieve the spectral phase of the pulse in a robust and precise way without the need of further diagnostic tools. The alignment is very easy compared to other techniques (no beam-splitting at any point, and no interferometric precision or stability are needed). This method is also particularly relaxed with respect to the necessary bandwidth of the SHG process, so relatively thick (tens of micrometers) frequency doubling crystals can be employed even when measuring few-cycle pulses.

## 2. Method

Consider an ultrashort laser pulse, which can be described by its complex spectral amplitude

$$\tilde{U}(\omega) = |\tilde{U}(\omega)| \exp\{i\phi(\omega)\} \quad (1)$$

The pulse goes through a piece of transparent glass and then a SHG crystal, and the measured SHG spectral power as a function of thickness is proportional to

$$S(\omega, z) = \left| \int \left( \int \tilde{U}(\Omega) \exp\{izk(\Omega)\} \exp(i\Omega t) d\Omega \right)^2 \exp(-i\omega t) dt \right|^2 \quad (2)$$

where $z$ is the thickness of the glass and $k(\Omega)$ its frequency-dependent phase per unit length (or wavenumber). Here, we simply take the original spectrum (amplitude and phase), apply a phase, and Fourier transform it to have the electric field in the time domain. Then SHG is performed (the time-dependent field is squared), and an inverse Fourier transform gives us the SHG spectrum. We perform a dispersion scan (we will call it d-scan for short) on the unknown pulse by introducing different thicknesses of glass and measuring the corresponding SHG spectra, which results in a two-dimensional trace. This is analogous to a MIIPS trace, but in our case the phase function is simply the one introduced by a piece of glass.

This model assumes that the SHG process consists simply on squaring the electric field in time, which assumes an instantaneous and wavelength-independent nonlinearity. We will discuss the consequences of this approximation later. For simplicity, we will also use negative values for the glass insertion. While this is obviously unrealistic from an experimental point of view, mathematically it simply results from setting a given reference insertion as zero. Regardless of this definition, if we know the electric field for a given insertion, it will be straightforward to calculate it for any other insertion.

As an example, we show in Fig. 1 calculated dispersion-scanned SHG traces of some representative spectra, where the spectral phase (left) refers to zero insertion in the d-scans (right). In all cases we used the same power spectrum, which is an actual spectrum measured from the few-cycle ultrafast oscillator used in the next section, and applied different phase curves. The assumed glass is BK7, and the corresponding phase was calculated from easily available Sellmeier equations.

The question now arises on how to find the electric field that generated a given scan. In MIIPS it consists on, for each wavelength $\lambda_0$ in the SHG spectrum, finding the insertion that maximizes this signal, noting the corresponding GDD at that point, and assuming that the GDD at the corresponding wavelength $2\lambda_0$ in the fundamental spectrum is the negative of this value. This works well for slowly varying phases, such as pure GDD and/or third-order dispersion (TOD), but fails for complex and structured phases. While the SHG at a given wavelength is mostly determined by the spectral power and phase at twice that wavelength in the fundamental field, there is always a coupling between all the generating and generated wavelengths.

We used this coupling to our advantage: by using the whole trace's information, combined with a numerical iterative algorithm, we are able to retrieve the spectral phase in a robust and precise way.

The method we used to retrieve the phase, although certainly not the only possible one, proved to be extremely flexible and reliable. It is based on the Nelder–Mead [13] (or downhill simplex) algorithm. We use the measured spectral power density, and by applying different phase curves, try to minimize a merit function (the rms error between the measured and simulated scans, as commonly used in FROG retrievals), given by

$$G = \sqrt{\frac{1}{N_i N_j} \sum_{i,j} \left(S_{meas}(\omega_i, z_j) - \mu S_{sim}(\omega_i, z_j)\right)^2} \qquad (3)$$

where $S_{meas}$ and $S_{sim}$ refer to the measured and simulated scans, respectively, and $\mu$ is the factor that minimizes the error. This factor, which can be easily found by differentiating the error with respect to $\mu$, is given by

$$\mu = \sum_{i,j} \frac{S_{meas}(\omega_i, z_j) S_{sim}(\omega_i, z_j)}{S_{sim}(\omega_i, z_j)^2}, \qquad (4)$$

and must be updated at each iteration. The problem can now be treated as a general multi-dimension optimization problem, where the phase is defined by a function of a set of parameters (or dimensions) and the function to be minimized is the error $G$. To make things easier for the algorithm, the phase function should be described in a convenient basis. We want to minimize the number of dimensions in the problem while still accurately describing the phase, and we want a basis whose functions are as uncoupled as possible, to prevent the algorithm from getting stuck on local minima. Different approaches can be taken here. Some authors choose to allow each point of the sampled complex spectral or time amplitude to be an independent variable (e.g. [14]), and as such, the number of dimensions of the problem will be determined by the sampling. Another (very common) choice is to use a Taylor expansion as a basis. In the former case, the large number of parameters makes the algorithm rather slow, while in the latter, there is a high degree of coupling between the even terms (i.e., second order dispersion, fourth order dispersion, etc.) as well as between the odd terms (third order dispersion, fifth order dispersion, etc.). This would still be a good choice (if not optimal) for simple phase functions, as the ones introduced by glasses, gratings, prism compressors, etc., which are accurately described in such a way.

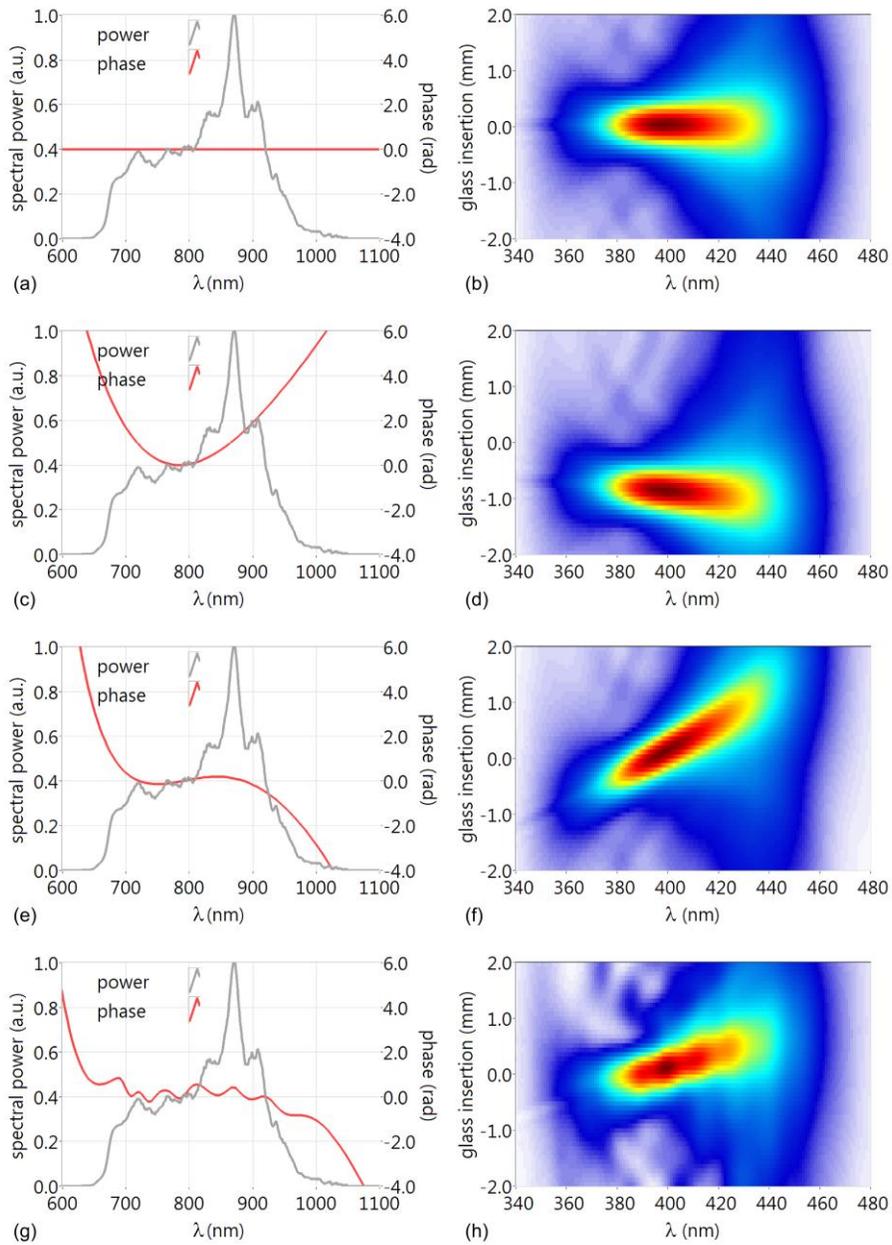

Fig. 1. Example of simulated dispersion scans, where the spectral phase plots on the left correspond to zero insertion in the scans on the right. (a) Fourier limited pulse. (b) Linearly chirped pulse (second-order dispersion only) – this causes mostly a translation of the trace with respect to the glass insertion, but since the glass itself doesn't introduce pure second order dispersion, the pulse is never completely compressed for any insertion, so it appears slightly tilted. (c) Pulse with third-order dispersion only, around 800 nm, which results in a clear tilt in the trace with respect to the previous cases. (d) A more complex phase curve, mostly third-order dispersion and some phase ringing.

In our case, we chose to write the phase as a Fourier series. This was inspired by the fact that Fourier components are orthogonal. If one could access directly the error between the true phase and its Fourier representation, then each Fourier component could be directly determined by minimizing the error. While we don't have direct access to this error, the overall trace error is a good indicator of the phase error. In fact, for all the cases we tried, the algorithm converged very well. For simple phases (i.e. mostly GDD and TOD) about 6 to 10 coefficients were used, whereas for more complicated phases up to 60 coefficients were used. The highest phase frequencies present on the fundamental spectrum can be estimated from the structure of its dispersion scan.

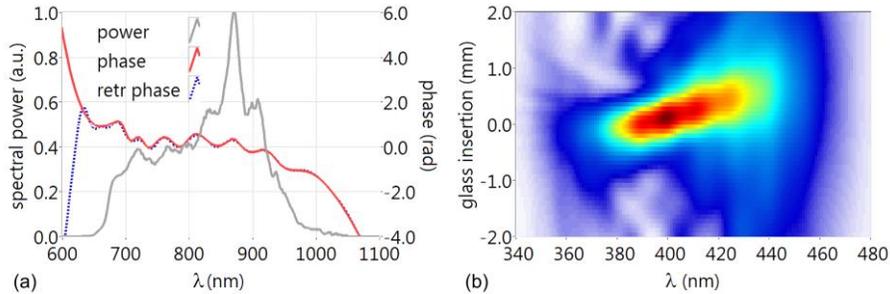

Fig. 2. Example of scan and phase retrievals from Fig. 1(h).

Fig. 2 shows an example of a simulated spectrum (measured power spectrum and simulated phase), its d-scan, and the corresponding retrieved phase. The agreement between the retrieved and original phases is very good typically down to regions where the spectral power is around 2% of the peak spectral power.

Let us now consider a more realistic scenario of particular importance for the case of ultra-broadband few-cycle pulses, where the SHG signal cannot be described by simply squaring the electric field (the SHG process doesn't have infinite bandwidth). Even in this case, the SHG signal is well described by the simple model (Eq. 2), provided the spectrum is multiplied by an adequate spectral filter [15,16], so the measured signal is simply given by

$$S_{meas}(\omega, z) = S_{ideal}(\omega, z) R(\omega) , \quad (5)$$

where $R(\omega)$ is the spectral filter and $S_{ideal}$ denotes the ideal, flat response process (Eq. 2). If the spectrometer's response to the SH signal is unknown it can also be included in this response function.

For the discussed algorithm, it is crucial to have a well calibrated signal, the reason being that the algorithm uses the overall error as a merit function. If the spectral response is not flat, the algorithm reacts by introducing fast phase variations on the regions with lower filter response, which makes the signal go out of the calculation box, therefore artificially reducing the overall error. There are several ways around this. The most straightforward would be to measure the spectrometer's response and simulate the SHG crystal spectral curve, but both are unfortunately difficult to obtain accurately. We found numerically that the integral of the trace over the thickness parameter (the frequency marginal)

$$M(\omega) = \int_{-\infty}^{+\infty} S(\omega, z) dz \quad (6)$$

does not depend on the original spectral phase of the pulse, $\phi(\omega)$. It is then easy to simulate a trace for a Fourier-limited pulse, and use its marginal to calibrate the measured one. Comparing the simulated scan's marginal to the measured scan's marginal it is straightforward to calculate the spectral response $R(\omega)$. Knowing the filter response, we can either divide the experimental trace by it, or include it in the retrieval process, by multiplying it by the "ideal" simulated trace, in each iteration. If the filter has zeros in the spectral region of interest, then we are left only with the latter option. We have successfully calibrated experimental scans this way.

We also devised another approach, which proved to be much easier to implement and more flexible. It consists in allowing the error function to be minimized for each wavelength, with the overall error being a weighted function of all these errors. So, given an experimental and simulated scan, the factor that minimizes the error for each frequency component is given by

$$\mu_i = \sum_j \frac{S_{meas}(\omega_i, z_j) S_{sim}(\omega_i, z_j)}{S_{sim}(\omega_i, z_j)^2} \quad (7)$$

and the overall error is

$$G = \sqrt{\frac{1}{N_i N_j} \sum_{i,j} \left( S_{meas}(\omega_i, z_j) - \mu_i S_{sim}(\omega_i, z_j) \right)^2} . \quad (8)$$

Now, by using this new error function, the algorithm effectively works on matching the trace's features, instead of simply trying to match the trace as a whole. If the trace is successfully retrieved, then the minimizing factors $\mu_i$ give us the complete filter response. What is perhaps more remarkable with this approach is that it is possible to correctly retrieve the phase for a certain frequency, even if there is no signal at the corresponding SHG (doubled) frequency. This can be seen from the examples in Fig. 3. Even in the case where the simulated filter response is clipped to zero (therefore making it impossible to calibrate the signal), the phase is nevertheless correctly retrieved across the whole spectrum. This would not be possible with the MIIPS retrieval technique.

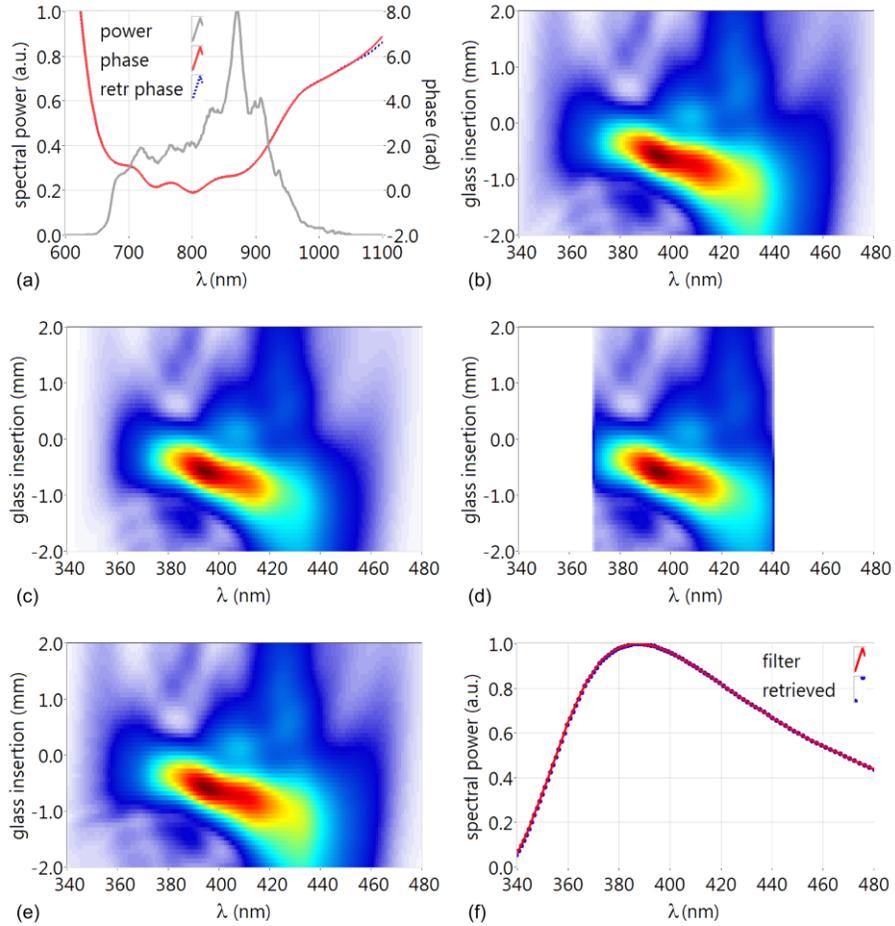

Fig. 3. Example of simulated traces including spectral filters in the SHG process. (a) Simulated spectrum, where the retrieved phase shown is for the worst case scenario, (d). (b) Ideal trace. (c) Ideal trace multiplied by a typical SHG crystal efficiency curve. (d) Same as (c), but clipped at around 370nm and 440nm. (e) Retrieved "ideal" scan from scan (d) – the retrieved scan is supposed to be identical to scan (b). (f) Applied and retrieved spectral filters from (c). The retrieved filter is made up of the error minimizing coefficients μ's for each wavelength.

## 3. Experimental results

A simplified diagram of our experimental setup is given in Fig. 4. It consists on an ultrafast oscillator (Femtolasers Rainbow CEP, not shown), four double-chirped mirror pairs (Venteon GmbH), followed by BK7 AR-coated glass wedges with an 8º angle, an off-axis aluminum-coated parabola (50mm focal length) and a standard 20 μm thick BBO crystal cut for type I SHG at 800 nm.

A dispersion scan was performed with very fine sampling in thickness (250 acquired spectra, with a thickness step of about 20 μm). Because of the relatively small angle of the wedges, this thickness step corresponds to a wedge translation step of more than 100 μm (and even this is much more than necessary, as a thickness step of 100 μm is typically enough, which corresponds to a translation step of more than 500 μm) so the positioning precision is quite undemanding compared to interferometric methods.

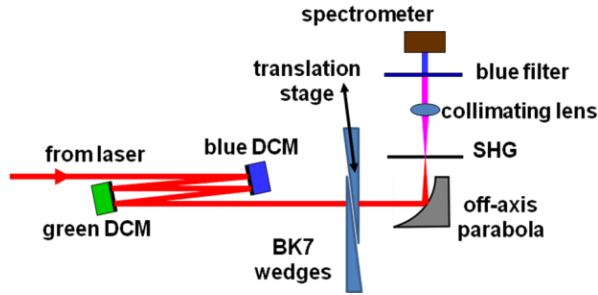

Fig. 4. Experimental setup. The laser is a Femtolasers Rainbow CEP (80 MHz repetition rate, energy per pulse of 2.5 nJ, FWHM Fourier limit of 6fs), SHG is a 20 μm thick BBO crystal. The double chirped mirrors (DCM) are made in matched pairs to minimize phase ringing, and the aluminum off-axis parabola has a 50 mm focal length.

To test the precision of the method, a bootstrap analysis was performed: from this fine scan, five scans were extracted, all with different datasets, by using every fifth spectrum (i.e., scan 1 uses steps 1,6,11, etc., scan 2 uses steps 2, 7, 12, etc.). The background signal was subtracted, and when the resulting signal was negative, we kept it as such, instead of making it zero. This way we allow for the retrieved data to (correctly) tend to zero where it should, instead of forcing the algorithm to try to converge to half of the noise level.

Two different retrieval techniques were used for each scan thus yielding a total of ten retrievals. In the first case we calibrated the scan from its frequency marginal (i.e., by forcing the integral over $z$ to be the same for the measured scan and for a simulated scan corresponding to the Fourier limit case), and in the second, we allowed the error to adjust to each spectral slice.

In all cases, the retrievals are very similar so we grouped them all together for the statistical analysis (Fig. 5). The "zero" insertion here refers to the insertion at which the pulse is shortest, and for which the phase and time reconstructions are shown. It actually corresponds to about 3 mm of BK7 glass. The retrieved pulse width was $7.1 \pm 0.1$ fs. The pulses clearly show the effect of residual uncompensated third order dispersion (also evidenced by the tilt in the corresponding d-scan trace) in the form of post-pulses. Note that there is no time-direction ambiguity on the retrieved pulse. Even if the laser and setup as it is don't allow for any shorter pulses, the precise phase measure allows one to re-design the compressor if necessary, i.e. by using different glasses and/or chirped mirrors.

It is worth noting that the phase retrieval is very robust even in regions of very low spectral power density. And, considering there is very little SHG signal above 470 nm and below 350 nm, it is surprising at first that the phase is consistently retrieved well beyond 940 nm and below 700 nm. Again, this is due to the coupling between all the frequency components on the trace and the original spectrum. As with FROG, the key aspect of this technique is the data redundancy in the dispersion-scanned SHG trace.

As with the simulated scans, it was possible to fully retrieve the filter response of the system as well. With both methods we retrieved very similar curves for all traces.

The phase retrieval technique used in this work is certainly not the only possible one. Even if it worked extremely well for our purposes, better, faster and more elegant numerical approaches are certainly possible and will be studied in future work.

Another advantage of using a multi-dimensional minimization technique is its extreme flexibility. For example, we tried feeding the algorithm the glass thickness spacing as a parameter, and it correctly found the known experimental value.

After having the field well characterized for a given insertion it is straightforward to calculate it for any other insertion by applying the known phase curve of the glass to the retrieved phase. One can then simply find the insertion that minimized the pulse length and move the wedges into the corresponding position.

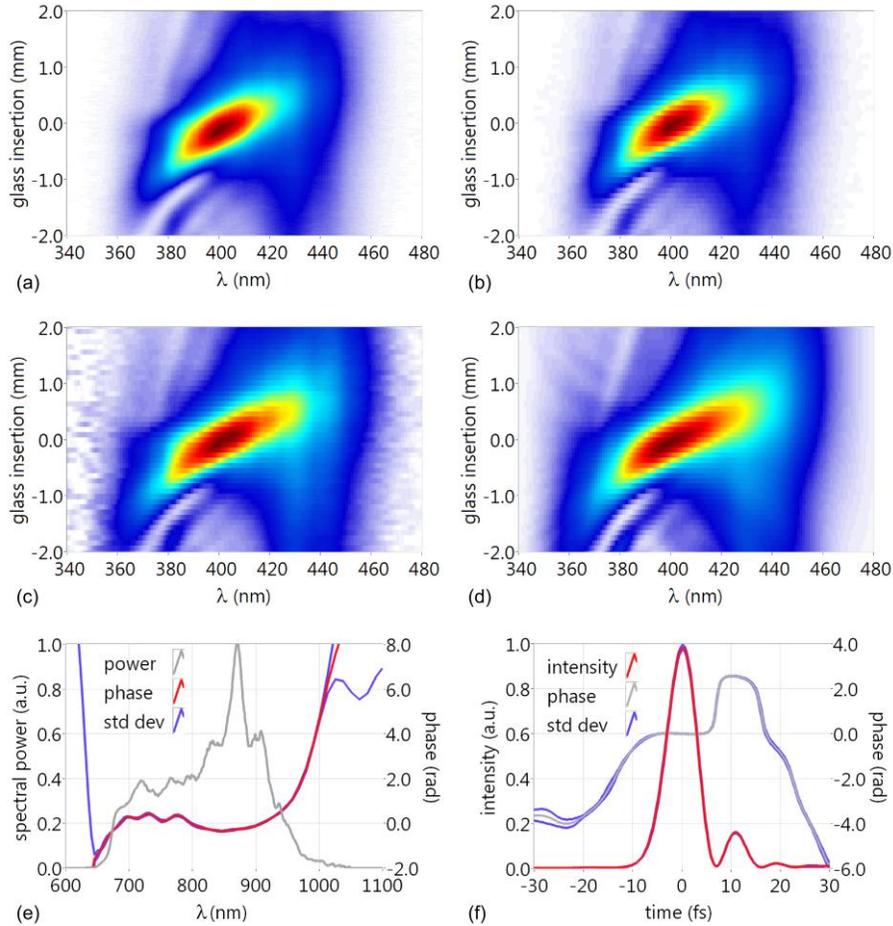

Fig. 5. Measured and retrieved scans. (a) Raw scan, made up of 250 spectra. (b) Scan made from 50 spectra out of the raw scan. (c) Calibrated scan, by using the frequency marginals in Eq. 6. (d) Retrieved scan from (c) - either retrieving from (c) or (b), the results are very similar. Plots (e) and (f) show a bootstrap analysis on spectrum and time, from 10 different retrievals. From the original scan with 250 spectra, 5 different scans were obtained using different datasets. The two different techniques were used on each dataset. The red curve is the average value, and the blue curves are one standard deviation above and below the average. Retrieved pulse width at FWHM was $7.1 \pm 0.1$ fs.

## 4. Conclusion

We have described and demonstrated a simple, inexpensive and robust method to characterize ultrashort laser pulses based on iterative phase retrieval from dispersion scans, using chirped mirrors, wedges and a standard (relatively thick) SHG crystal. The alignment is very easy (no beam-splitting at any point, and no interferometric precision or stability are needed). In our case, the main part of the setup (chirped mirrors and wedges) was already being used for pulse compression, so there was no need to employ other characterization methods. This is the situation where this technique is especially useful. It is of course possible to use the system as a standalone device. Also, we are not as limited by the phase-matching restrictions of the SHG crystal as with other techniques, which allows for the characterization of extremely broad bandwidth pulses without having to sacrifice SHG efficiency by employing unpractically thin crystals. We believe this technique might be immediately useful for many people working in the field with pulse compressors based on chirped mirrors.

**Acknowledgements**

This work was partly supported by FCT – Fundação para a Ciência e a Tecnologia and FEDER (grants SFRH/BD/37100/2007 and PTDC/FIS/115102/2009)， the European Research Council (ALMA), the Marie Curie Intra-European Fellowship ATTOCO, the Marie Curie Initial Training Network ATTOFEL, the Knut and Alice Wallenberg Foundation, the Joint Research Programme ALADIN of Laserlab-Europe II and the Swedish Research Council.